\documentclass[prb,twocolumn,showpacs,preprintnumbers,amsmath,amssymb]{revtex4}
\usepackage{graphicx}
\usepackage{epsfig}
\newcommand{\be}{\begin{eqnarray}}
\newcommand{\ee}{\end{eqnarray}}

\newcommand{\bi}{\begin{itemize}}
\newcommand{\ei}{\end{itemize}}

\begin{document}
\title{Boundary condition for Ginzburg-Landau theory 
         of superconducting layers}

\author{Jan Kol\'a\v{c}ek$^1$
      , Pavel Lipavsk\'y$^{1,2}$
      , Klaus Morawetz$^{3,4}$ 
    and Ernst Helmut Brandt$^5$}
\affiliation{
$^1$Institute of Physics, Academy of Sciences, 
Cukrovarnick\'a 10, 16253 Prague 6, Czech Republic}
\affiliation{
$^2$ Faculty of Mathematics and Physics, Charles University, 
Ke Karlovu 3, 12116 Prague 2, Czech Republic}
\affiliation{
$^3$Forschungszentrum Dresden-Rossendorf, PF 51 01 19, 
01314 Dresden, 
Germany}
\affiliation{$^4$ International Center for Condensed Matter Physics, 
70904-910, Bras\'ilia-DF, Brazil}
\affiliation{
$^5$Max Planck Institute for Metals Research,
         D-70506 Stuttgart, Germany}
\begin{abstract}
Electrostatic charging changes the critical temperature of superconducting
thin layers. To understand the basic mechanism, it is possible to use the
Ginzburg-Landau theory with the boundary condition derived by 
de Gennes from the BCS theory. Here we show that a similar boundary 
condition can be obtained from the principle of minimum free energy. 
We compare the two boundary conditions and use the Budd-Vannimenus 
theorem as a test of approximations.
\end{abstract}
	
\maketitle

\section{Introduction}
Much experimental effort is devoted to find 
superconducting materials with critical temperatures as high 
as possible. It is well known that the critical temperature depends 
on the charge carrier density. 
The charge carrier density can be changed by doping and to some extend 
it can also be changed by electrostatic charging. Consequently it is an attractive 
task to determine, how electrostatic charging evoked by an applied
electric field changes the critical temperature 
of superconductors. Generally the experiments revealed 
that it is easier to increase $T_{\rm c}$ than to decrease it \cite{MGT03, 
Matijasevic94}. The Ginzburg-Landau (GL) theory with the de Gennes boundary condition 
can be used to understand this behavior \cite{LMKY06}. 

The aim of this paper is to show
how a superconductor screens the external electric field and
to which extent the boundary condition derived from the minimum free energy
principle is compatible with the de Gennes boundary condition. 

Charges at a solid surface partially leak out of the surface. 
This creates a surface dipole. The Budd-Vannimenus theorem 
\cite{04BernoulliDipole} describes the step in the surface potential due 
to this surface dipole as a simple expression 
of the bulk free energy density. Therefore it is well suited to test 
the approximations used in this paper. 

In the first chapter we explain the model and the parts considered in the free energy 
of the superconductor and solve the Euler-Lagrange equations for the GL and charge carrier 
wave function and the surface potential. In chapter III the corresponding equations 
outside the superconductor are solved and in chapter IV the continuity requirements determine 
the remaining constants. Chapter V presents the numerical values which are compared 
with the de Gennes boundary condition in chapter VI. Finally we conclude in chapter VII. 
 
\section{Free energy in the superconductor}
We start with the free energy 
\begin{equation}
{\cal F}=\int \left( f_{\rm TF}+f_{\rm GC}
             +f_{\rm GL}+f_{\rm elst} \right)d{\bf r},
\label{FreeEn}
\end{equation}
where we include only the terms most relevant for the above specified problem. 

The first term $f_{\rm{TF}}$ is the Thomas-Fermi 
internal energy, for which we use the LDA (local-density approximation) 
\begin{equation}
f_{\rm TF}=\dfrac{3}{5} \left( 3 \pi^{2} \right)^{\frac{2}{3}}
       \frac{\hbar^2}{2m} n^{\frac{5}{3}}.
\label{fTF}
\end{equation}
The second term $f_{\rm {GC}}$ represents the condensation energy 
for which we use the formula following from the Gorter-Casimir 
two fluid model \cite{GC34,GC34b}
\begin{equation}
f_{\rm GC}=
          \frac{1}{4}
          \gamma T_c^2
          \left(\frac{n_s}{n}+2\frac{T^2}{T_c^2}\sqrt{1-\frac{n_s}{n}}\right).
\label{fGC}
\end{equation}
The electrostatic energy density term reads
\begin{equation}
f_{\rm elst}=
    \frac{1}{2}
    \epsilon_0{\bf E}^2 + e \varphi \delta n,
\label{felst}
\end{equation}
in the form suitable for performing variations.
For simplicity we exclude the magnetic field and it's related 
kinetic energy of the screening current. We take the 
vector potential $\bf A$ to be zero and write the GL gradient term as
\begin{equation}
f_{\rm GL}=
          \frac{\hbar^2}{2m^*}
          \psi_n^2\left| \nabla\psi\right|^2.
\label{fGL}
\end{equation}
Here we have chosen the GL wave function $\psi$ normalized with respect 
to the total charge carrier density $n$. In the spirit of the
Thomas-Fermi approximation the charge carriers are described 
by a wave function $\psi_n$ with $n=\psi_n^2$ and the 
superconducting fluid density used in the formula for the
condensation energy (\ref{fGC}) reads 
\begin{equation}
n_s=2\psi_n^2\psi^2.
\label{dfns}
\end{equation}

In short, the free energy is expressed by three independent variables: 
the scalar potential $\varphi$ determining the electric field 
${\bf E} = -\nabla \varphi$, 
the GL wave function $\psi$ and the charge carrier
wave function $\psi_{\rm n}$. 
We assume that the material parameters, 
the critical temperature $T_c$ and 
the Sommerfeld parameter $\gamma$ depend on the charge carrier's density $n$, 
by using the approximations 
\begin{eqnarray}
\gamma\left(n\right)
         &=&\gamma_0 \left(1+\frac{n-n_{\rm lat}}{n_{\rm lat}} 
         \frac{\partial{\rm ln} \gamma}{\partial {\rm ln}  n} \right),\\
\label{dfgaman}
T_c\left(n\right)&=&T_{\rm c0} \left(1+\frac{n-n_{\rm lat}}{n_{\rm lat}} 
         \frac{\partial{\rm ln} T_{\rm c}}{\partial {\rm ln}  n} \right),
\label{dfTcn}
\end{eqnarray}
where $n_{\rm lat}$ is the crystal lattice density.
In the following we shall use 
three characteristic length:\\
(i) the Thomas-Fermi screening length
\large 
$\lambda_{\rm TF}^2=\frac{2\epsilon_0}{3ne^2}E_{\rm F}$,\\
\normalsize
(ii) the Bohr radius
\large 
$a_{\rm B}=\frac{4\pi\epsilon_0\hbar^2}{me^2}$\\
\normalsize
(iii) and the coherence length
\large
$\xi_0^2=\frac{\hbar^2n_{\rm lat}}{4\gamma_0T_{\rm c0}^2 m}.$
\normalsize

From the charge neutrality requirement we know that 
$\psi_{{\rm n}\infty}=\sqrt{n_{\rm lat}}$  
(here and in the following the subscript $\infty$ denotes
the magnitude far from the surface).
To keep things simple, we use the approximations 
\begin{eqnarray}
\psi_{\rm n}&=&\sqrt{n_{\rm lat}}\left(1+\delta\psi_{\rm n} \right)\\
\label{aprpsin}
\psi&=&\psi_\infty \left(1+\delta\psi \right)\\
\label{aprpsi} 
e\varphi&=&E_{\rm F}\left(\varphi_\infty+\delta\varphi \right)
\label{aprphi}
\end{eqnarray}
and suppose that the deviations of the three independent 
variables from the optimum values are small. 
In a homogeneous superconductor far from the surface all these deviations  
have a zero value and the derivatives 
of the free energy formula (\ref{FreeEn}) with respect to them must be also zero.
From these requirements we get the magnitudes of the optimum superfluid fraction  
\begin{equation}
\frac{\psi_\infty}{\psi_{{\rm n}\infty}} = \sqrt{1-t^4}
\label{dfpsiinf}
\end{equation}
and the optimum magnitude of the scalar potential 
\begin{equation}
{\varphi_\infty}
=-1+\frac{2\lambda_{\rm TF}^4}{\pi^2 a_{\rm B}^2\xi_0^2}
\left( 2 \left(1-t^4 \right)\frac{\partial {\rm ln}T_{\rm_c}}
                                  {\partial {\rm ln}n }
+\left(1+t^4 \right)\frac{\partial {\rm ln}\gamma}
                                  {\partial {\rm ln}n }\right).
\label{dfphiinf}
\end{equation}
The electrostatic potential energy of the charge carrier thus equals the 
Fermi energy 
\begin{equation}
E_{\rm F}= \frac{\hbar^2}{2m}\left(3\pi^2 n \right)^{2/3},
\label{dfEF}
\end{equation}
with a small (lower than gap) correction represented 
by the second term in (\ref{dfphiinf}). 

Using the second order expansion of the free energy (\ref{FreeEn}), 
from the variation we get three linear 
Euler-Lagrange (EL) equations for the three independent variables,
\begin{equation}
\frac 3 4\lambda_{\rm TF}^2 \nabla^2 \delta\varphi+\delta\psi_{\rm n} =0,
\label{ELphi}
\end{equation}

\begin{equation}
   \left(1-t^4\right)\xi_t^2\nabla^2\delta\psi
  +t^4\delta\psi+4t^4
  \frac{\partial {\rm ln}T_{\rm c}}{\partial{\rm ln}n}
  \delta\psi=0,  
\label{ELpsi}
\end{equation}

\begin{eqnarray}
\left(2t^4\frac{\partial {\rm ln}T_{\rm c}}{\partial{\rm ln}n}
  \left(\frac{\partial {\rm ln}T_{\rm c}}{\partial{\rm ln}n}
    +2\frac{\partial {\rm ln}\gamma}{\partial{\rm ln}n}
  \right) 
    -\frac{\pi^2a_{\rm B}^2\xi_t^2}{12\lambda_{\rm TF}^4}
 \right)\delta\psi_{\rm n}\nonumber \\
+2t^4\frac{\partial {\rm ln}T_{\rm c}}{\partial{\rm ln}n}\delta\psi
    -\frac{\pi^2a_{\rm B}^2\xi_t^2}{16\lambda_{\rm TF}^4}\delta\varphi
=0. 
\label{ELpsin}
\end{eqnarray}
Close to the planar surface we can assume exponential dependencies 
of the deviations and from the EL equations (\ref{ELphi}-\ref{ELpsin})
we get a second order equation for the square of the expected 
penetration depth $\lambda$. Two solutions arise out of it.

Observing that $\lambda_{\rm TF}, a_{\rm B}\ll\xi_0$, we find a 
first approximate solution $\lambda$ in the form of the coherence-like length 
\begin{equation}
\xi_t=\xi_0 \frac{2t^2}{\sqrt{1-t^4}}.
\label{dfxit}
\end{equation}
In this solution the scalar potential is constant 
and local charge neutrality is preserved.
Only the deviation of the wave function $\delta\psi$ is nonzero 
($C_\xi$ will denote its magnitude).

A very small penetration depth characterizes the second solution such that 
this solution can be simplified.  
Using the same approximation as above, we find that the second penetration depth
equals the Thomas-Fermi screening length
$\lambda_{\rm TF}$. In this solution the scalar potential displays a sharp step 
($C_{\rm TF}$ will denote its magnitude) and from the Poisson equation 
(\ref {ELphi}) follows that the charge carrier's 
density changes accordingly.
The sharp step on the GL wave function $\psi$ is negligibly small
due to the factor $\lambda_{\rm TF}^2/\xi_t^2$ which enters
the resulting formula. 
It corresponds to the well known fact that the GL wave function 
$\psi$ cannot abruptly change. 

The general solution can be written as a sum of the two above described 
solutions:
\begin{eqnarray}
\delta\varphi&= &C_{\rm TF} {\rm exp}\left(\frac{-x}{\lambda_{\rm TF}}\right)
\label{soldelphi} \\
\delta\psi_{\rm n}&= &-\frac{3}{4}C_{\rm TF} {\rm exp}       
      \left(\frac{-x}{\lambda_{\rm TF}}\right)\\
\label{soldelpsin}
\delta\psi&= &\frac{3\lambda_{\rm TF}^2 t^4}{(1-t^4)\xi_t^2}
           \frac{\partial{\rm ln}T_{\rm_c}}{\partial {\rm ln}n}
             C_{\rm TF} {\rm exp}       
                  \left(\frac{-x}{\lambda_{\rm TF}}\right)
\nonumber \\
              && +C_\xi {\rm exp}
                 \left( \frac{-x}{\xi_t}\right).
\label{soldelpsi}
\end{eqnarray}
Here  $C_{\rm TF}$ describes the step of the scalar potential in units of $E_{\rm F}/e$ according to (\ref{aprphi}).
Using (\ref{soldelphi}) - (\ref{soldelpsi}) we can calculate the free energy
\begin{eqnarray}
{\cal F}=\int
  \left( \frac{3}{5}+\frac{2(1+t^4)\lambda_{\rm TF}^4}
                         {\pi^2 a_{\rm B}^2\xi_0^2}
  \right) {\rm d}{\bf r} 
  -\frac{3}{4}\lambda_{\rm TF} C_{\rm TF}^2
  \nonumber\\ 
  +\frac{8 \lambda_{\rm TF}^4 (1-t^4)}{\pi a_{\rm B}^2 \xi_t} C_\xi^2
  +\frac{48 t^4 \lambda_{\rm TF}^5 }{\pi^2 a_{\rm B}^2 \xi_t^2} 
   \frac{\partial{\rm ln} T_{\rm c}}{\partial {\rm ln}  n} 
  C_{\rm TF}C_\xi  .
\label{Fgensol}
\end{eqnarray}
For a semi-infinite medium the first term gives an
infinite contribution which is not influenced by the surface 
conditions, so we do not need to deal with it. 
The last three terms correspond to the surface energy, 
which according to the principle of minimum free energy 
should take an extremum. 
The minimum of the free energy is obtained for
\begin{equation}
C_\xi=-\frac{3\lambda_{\rm TF}\xi_t}{4\xi_0^2} 
        \frac{\partial{\rm ln} T_{\rm c}}{\partial {\rm ln}  n}
        C_{\rm TF},
\label{Cximin}
\end{equation}
in which case the derivative of the wave function $\psi$ at the surface is zero. 
For lead at $t$ = 0.9 we get
$C_\xi = - 0.00044 C_{\rm TF}$. As expected,  
the deviation of the GL wave function $\delta \psi$ is much smaller
compared to the sharp steps on the scalar potential $\delta \varphi$ 
and on the charge carrier wave function $\delta \psi_{\rm n}$. 

We see that the principle of minimum free energy entails the 
GL boundary condition. 
Towards the surface the GL wave function $\delta \psi$ displays
a small gradual change, only very close to the surface its derivative 
jumps to zero. The solution is complete, if the parameter $C_{\rm TF}$ 
is determined. It can be derived from the requirement 
of continuity with a solution minimizing the total
free energy including the one of the vacuum outside.

\section{Free energy outside the slab}
Now we approximate the free energy density outside the superconductor by
\begin{equation}
{\cal F}=\int \left( f_{\rm W}
             +f_{\rm GL}+f_{\rm elst} \right)d{\bf r}.
\label{Fout}
\end{equation}
We include the electrostatic term, the GL gradient correction and the
von Weizs\"acker kinetic energy functional
\begin{equation}
f_{\rm W}=\frac{\hbar^2}{2m}\left| \nabla\psi_{\rm n} \right|^2.
\label{dffW}
\end{equation}
In the limit of rapidly varying densities this kinetic energy term is 
dominant and when describing charge carriers tunneling outside 
the superconductor this term cannot be omitted. 
We have not included this term into the formula (\ref{fGL}) describing the free energy 
inside. The reason is that inside the superconductor 
the Thomas-Fermi internal energy plays the dominant role and moreover, as it is 
shown e.g. in the book of Dreizler and Gross \cite{Drez90}, in the limit of nearly 
homogeneous systems the second order term of the gradient expansion provides a
better approximation. It has the same structure as the von Weizs\"acker kinetic 
energy functional, but its  coefficient is nine times lower. 
We suppose that for the rough estimates presented here this relatively small correction 
can be neglected. 

In the vacuum far from the surface the scalar potential reaches 
the magnitude of the work function $\varphi_{\rm W}$, so that 
we can approximate
\begin{equation}
e\varphi=E_{\rm F}\left( \varphi_{\rm W}+\delta\varphi \right). 
\label{aprphiout}
\end{equation}
The density of the tunneling charge carriers quickly drops to zero. Using an analogous notation as above we write 
\begin{equation}
\psi_{\rm n}=\sqrt{n_{\rm lat}} \delta\psi_{\rm n} 
\label{aprpsinout}
\end{equation}
supposedly that $\delta\psi_{\rm n}$ is small.
For the wave function $\psi$ we use the approximation
\begin{equation}
\psi=\widetilde{\psi}_\infty +\delta\psi,
\label{aprpsiout}
\end{equation}
where $\widetilde{\psi}_\infty$ represents the superfluid fraction 
in the vacuum far from the surface. 
Let us remind that $\psi$ is normalized to the charge carrier density, 
see (\ref {dfns}), so that $\widetilde{\psi}_\infty$  
does not need to be zero. 
The free energy density in the vacuum thus reads
\begin{eqnarray}
f_{\rm out}=
     \frac{8\lambda_{\rm TF}^4}
       {\pi^2 a_{\rm B}^2}
       \left(2\left(\nabla\delta\psi_{\rm n}\right)^2
       +\delta \psi_{\rm n}^2 \left( \nabla \psi \right)^2\right)
       \nonumber\\
    -\frac{3}{4}\lambda_{\rm TF}^2
         \left(\nabla\delta\varphi\right)^2
     +\left(\varphi_{\rm W} +\delta\varphi\right)\delta\psi_{\rm n}^2
\label{fout}
\end{eqnarray}
and we can write the Euler-Lagrange equations. 

The variation with respect to the wave function $\psi$ gives 
the condition 
\begin{equation}
\delta\psi_{\rm n}^2\nabla^2\delta\psi=0.
\label{ELpsiout}
\end{equation}
We see that $\delta \psi$ remains constant or changes linearly. 

The proximity effects indicate that the 
correlated charge carriers can remain correlated even 
if they are tunneling. 
For simplicity we suppose, that the superfluid 
fraction of the charge carriers tunneling outside the material 
does not change, so we take $\delta\psi=\psi(0)$.
The two other Euler-Lagrange equations read 
\begin{equation}
2\varphi_{\rm W}\delta\psi_{\rm n} 
   -\frac{32\lambda_{\rm TF}^4}{\pi^2 a_{\rm B}^2}
     \nabla^2\delta\psi_{\rm n}=0
\label{ELpsinout}
\end{equation}
\begin{equation}
\delta\psi_{\rm n}^2 
   +\frac{3}{2}\lambda_{\rm TF}^2\nabla^2 \delta\varphi=0.
\label{ELphiout}
\end{equation}
In the same way as above we can try the exponential solution 
\begin{eqnarray}
\delta\psi_{\rm n} = K_{\rm n}\exp\left( \frac{x}{\lambda_{\rm W}}\right) 
\\
\delta\varphi = K_\varphi\exp\left( \frac{2x}{\lambda_{\rm W}}\right),
\label{phipsiout}
\end{eqnarray}
where $\lambda_{\rm W}$ denotes the tunneling length which follows from the Euler-Lagrange equation (\ref{ELphiout}) as
\begin{equation}
\lambda_{\rm W} =\frac{\sqrt{-6K_\varphi}}{K_{\rm n}}.
\label{sollambdaW}
\end{equation}
The work function can be determined from (\ref{ELpsinout}) as 
\begin{equation}
\varphi_{\rm W}=\frac{16\lambda_{\rm TF}^4}
    {\pi^2 a_{\rm B}^2\lambda_{\rm W}^2}.
\label{solphiW}
\end{equation}
In this way we have an approximate solution outside the superconductor,
which should be linked to the solution inside.

\section{Continuity requirements}
At the surface the continuity of the wave function $\psi_{\rm n}$ and the
continuity of the scalar potential $\varphi$ with 
its derivative (continuity of the electric field) must be ensured. 
We get three conditions
\begin{eqnarray}
\frac{2 \lambda_{\rm TF}^4}{\pi^2 a_{\rm B}^2\xi_0^2}
   \left(2\left(1-t^4\right) \frac
    {\partial{\rm ln} T_{\rm c}}{\partial {\rm ln}  n}
  +    \left(1+t^4\right) \frac
    {\partial{\rm ln} \gamma}{\partial {\rm ln}  n}\right)
    \nonumber\\
  -1+C_{\rm TF}  
 = K_{\rm \varphi}-\frac{8\lambda_{\rm TF}^2}{3\pi^2 a_{\rm B}^2}
     \frac{K_{\rm n}^2}{K_\varphi}
\label{con_phi}
\end{eqnarray}
\begin{equation}
-C_{\rm TF}=\frac{2K_\varphi K_{\rm n}}{\sqrt{-6K_\varphi}}+E_{\rm a}
\label{con_E}
\end{equation}
\begin{equation}
1-\frac{3}{4} C_{\rm TF} = K_{\rm n}
\label{con_psin}
\end{equation}
where the term $E_{\rm a}$ representing the applied electric field 
is included into the condition of continuity for the electric field. 
From the continuity requirements (\ref{con_phi}-\ref{con_psin}) 
we obtain the equation 
\begin{eqnarray}
  \frac{2 \lambda_{\rm TF}^4}{\pi^2 a_{\rm B}^2\xi_0^2}
   \left(2\left(1-t^4\right) \frac
    {\partial{\rm ln} T_{\rm c}}{\partial {\rm ln}  n}
  +    \left(1+t^4\right) \frac
    {\partial{\rm ln} \gamma}{\partial {\rm ln}  n}\right)
    \nonumber\\
  -\frac{\lambda_{\rm TF}^2\left(-4+3C_{\rm TF}\right)^4}
        {144\pi^2 a_{\rm B}^2\left(C_{\rm TF}-E_{\rm a}\right)^2}
  +\frac{24\left(C_{\rm TF}-E_{\rm a}\right)^2}
        {\left(-4+3C_{\rm TF}\right)^2}
    \nonumber\\
         -1+C_{\rm TF} =0,
\label{eq_CTF}
\end{eqnarray}
determining the step of the scalar potential $C_{\rm TF}$.

\section{Numerical values}
The sixth order equation (\ref{eq_CTF}) can be numerically solved. 
For small applied electric fields the linear expansion
\begin{equation}
C_{\rm TF}=C_{\rm TF0}+\zeta E_{\rm a}
\label{TF_Ea}
\end{equation}
is applicable and for lead at temperature $t$=0.9 
we get the numerical solution
\begin{equation}
C_{\rm TF}=0.457 - 0.53 E_{\rm a} .
\label{numCTF}
\end{equation}
The numerical estimate for the tunneling length follows to be  
from (\ref{sollambdaW}) $\lambda_{\rm W}=3.17\; \lambda_{\rm TF}$
and the work function according to (\ref{solphiW}) $\varphi_{\rm W}$= 1.43 eV. 
Taking into account how many simplifications we have used, 
it is surprising that the obtained results seem to be quite reasonable. 
The estimated magnitude of the work function is comparable with the
experimentally determined value of
$\varphi_{\rm W}$= 4.25 eV. \cite{Michaelson77}

The sharp step of the scalar potential can be estimated from
the modified Budd-Vannimenus theorem \cite{04BernoulliDipole} 
according to which 
\begin{equation}
e \left( \varphi_\infty -\varphi_0 \right)
    = \left(\frac{\partial f_{\rm el}}{\partial n}-\frac{f_{\rm el}}{n}\right).
\label{BuddVann}
\end{equation}
Here $f_{\rm el}$ denotes the spatial density of the electronic free energy,
which can be roughly approximated by the Thomas-Fermi internal energy $f_{\rm TF}$ 
defined in (\ref{fTF}). Then 
the Budd-Vannimenus theorem  (\ref{BuddVann}) predicts a sharp step, $C_{\rm TF}=\frac{2}{5}$, of the 
scalar potential in units of $E_{\rm F}/e$. 
The numerical solution (\ref {numCTF}) gives 
a comparable result what strongly supports the applicability of the here 
used approximations.   

\begin{figure} [htb]   
\centerline{\parbox[c]{8cm}{
\psfig{figure=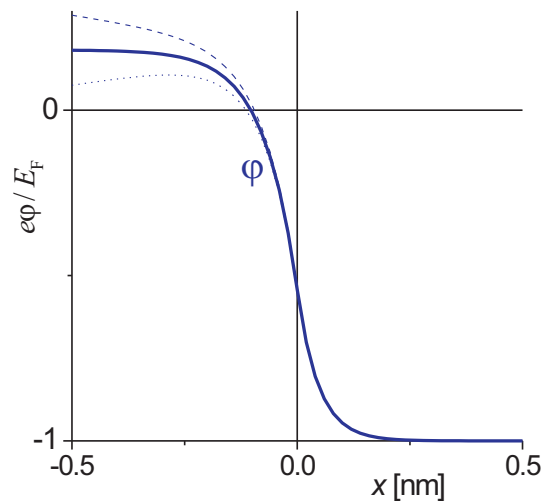,width=8cm}}}
\vskip -5mm
\caption{The electrostatic potential $\varphi$ calculated for
lead at $t=0.9 $K. The dashed and dotted lines correspond to applied 
electric field $E_a = \pm 0.01$} 
\label{figphi}
\end{figure}

\begin{figure} [tb]   
\centerline{\parbox[c]{8cm}{
\psfig{figure=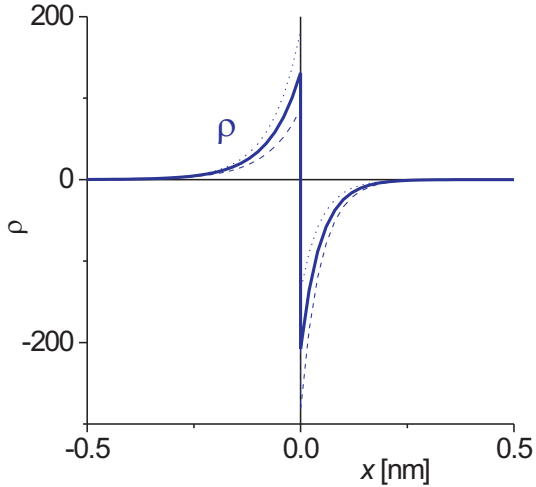,width=8cm}}}
\vskip -5mm
\caption{Charge density $\rho$. 
To make the screening visible, the dashed and dotted lines correspond 
to the applied electric field $E_a = \pm 0.3$} 
\label{figrho}
\end{figure}

\begin{figure} [htb]   
\centerline{\parbox[c]{8cm}{
\psfig{figure=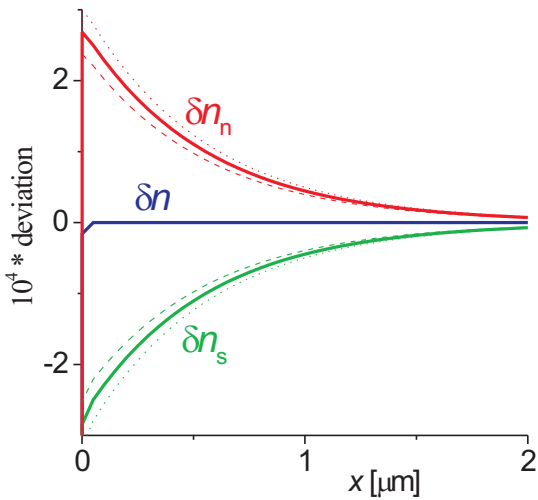,width=8cm}}}
\caption{Deviations of charge carrier densities. To ensure visibility,
the dashed and dotted lines correspond to the applied field
$E_a = \pm 0.1$.} 
\label{devCCdens}
\end{figure}

We saw that the numerical values of the measurable quantities 
are reasonable. 
In Fig. \ref{figphi} the scalar potential is plotted.
As expected, inside the superconductor the scalar potential 
acquires the Fermi energy value, while in the vacuum outside it reaches 
the work function value $\varphi_{\rm w}$. 
The dashed and dotted lines correspond to the experimentally 
accessible applied electric field 
$E_a = \pm 0.01 \frac{E_{\rm F}}{e \lambda_{\rm TF}}
       \cong \pm 1.7\times10^7$ V/cm. 
As it is seen in Fig. \ref{figrho}, the external electric field is 
screened on the Thomas-Fermi screening length.
In the figure \ref{devCCdens} deviations of the charge carrier densities 
from the equilibrium values are plotted. 
We can see that the magnitudes of these deviations are small. Close to the surface 
the superfluid density $n_{\rm s}$ decreases and this decrease is compensated 
by an increase of the normal fluid density $n_{\rm n}$. The total 
charge carrier density $n$ shows no change on the scale of the
coherence length.  

\section{Comparisons with de Gennes formula}
\begin{figure} [t]   
\centerline{\parbox[c]{8cm}{
\psfig{figure=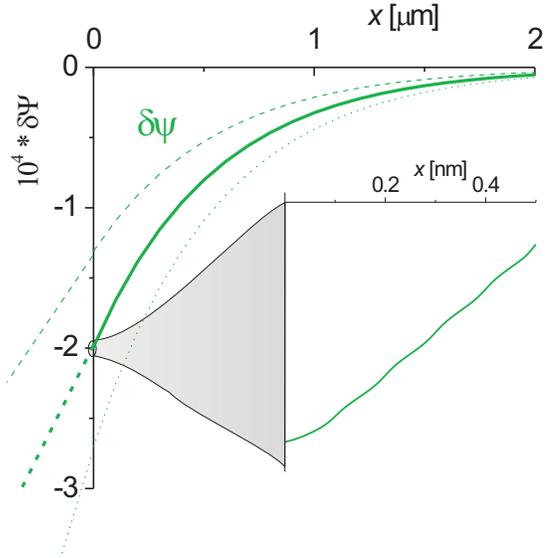,width=8cm}}}
\vskip -5mm
\caption{Deviation 
$\delta\psi$ of the wave function $\psi$ from the
equilibrium value. Close to the surface its derivative 
approaches zero (see the insert). The dashed and dotted lines indicate
how the extrapolation length changes with applied field.} 
\label{figpsi} 
\end{figure}

Now we compare the GL boundary condition following 
from the minimum free energy principle with 
the de Gennes boundary condition \cite{G66}
\begin{equation}
\left. \frac{\nabla\psi}{\psi} \right|_0
   =\left. \frac{\nabla\Delta}{\Delta} \right|_0
   =\frac{1}{b}
   =\frac{1}{b_0} +\frac{E_{\rm a}}{U_{\rm s}},
\label{deG}
\end{equation}
according to which the derivative of the gap at the surface 
is not exactly zero even without external electric field. 
The zero field extrapolation length $b_0$ is 
around 1 cm (almost infinity from the microscopic point of view). 
The effective potential $U_{\rm s}$
\begin{equation}
\frac{1}{U_{\rm s}}=\frac{3\eta\lambda_{\rm TF}^2}{2\xi_0^2}
       \frac{\partial{\rm ln}T_{\rm c}}{\partial{\rm ln}n}
       \frac{e}{E_{\rm F}}
\label{Us_dG}
\end{equation}
determines how the extrapolation length $b$ changes if an external electric field 
$E_{\rm a}$ is applied \cite{LMKY06}. 
De Gennes estimated the surface ratio 
\begin{equation}
\eta\equiv\frac{\Delta(0)}{\Delta_0}
\label{dfeta}
\end{equation}
to be close to one. 
For lead the formula (\ref{Us_dG}) gives $U_{\rm s}  = 1.35 \times 10^7$ V.
From the minimum free energy we know, however, that the derivative at the surface 
should be zero. In Fig.\ref{figpsi} we see how the deviation of the 
wave function $\psi$ at the surface decreases with the derivative determined 
by the parameter $C_\xi$. From this we get the extrapolation 
length $b_0\approx 2.8$ mm, a value comparable with the one estimated 
by de Gennes. Only very close to the surface (on the distance of Thomas-Fermi 
screening length) the derivative of the wave function $\psi$ approaches zero 
(see insert of the Fig. \ref{figpsi}).

In figure \ref{figpsi} we can also observe how the extrapolation length 
changes if an electric field $E_{\rm a}$ is applied. 
By substituting $b=-\xi_{\rm t}/C_\xi$ into (\ref{deG}) 
and using (\ref{Cximin}) with the approximation (\ref{TF_Ea}) we get 
a simple expression for the effective potential $U_{\rm s}$ 
\begin{equation}
\frac{1}{U_{\rm s}}=\frac{3\zeta\lambda_{\rm TF}^2}{4\xi_0^2}
       \frac{\partial{\rm ln}T_{\rm c}}{\partial{\rm ln}n}
       \frac{e}{E_{\rm F}}.
\label{Us_fE}
\end{equation}
This formula is similar to the de Gennes formula (\ref{Us_dG}). 
We should notice, however, that in this formula the extrapolation 
parameter $\zeta/2$ of (\ref{TF_Ea}) appears instead of the surface ratio $\eta$
which enters de Gennes formula (\ref{Us_dG}).

\section{Conclusions}
It was shown in this paper that the minimum free energy principle entails a
zero derivative of the wave function $\psi$ at the surface
of the superconductor. On the scale of the coherence length,
however, even if no external electric field is applied, 
the derivative is nonzero and its magnitude corresponds 
to the de Gennes estimate. Only on the Thomas-Fermi screening length 
scale it approaches zero. 
In the presence of an external electric field the extrapolation
length changes according to equation (\ref{deG}), with
the effective  potential given by equation (\ref{Us_fE}).
This formula is similar to formula (\ref{Us_dG}) following
from the de Gennes theory. 
The agreement with the Budd-Vannimenus theorem  and the numerical 
estimates support the applicability of the proposed
approach. 

\acknowledgments

This work was supported by the  Czech research plans MSM 0021620834 and 
No. AVOZ10100521, by grants GA\v{C}R 202/07/0597 and 202/08/0326 
and GAAV IAA100100712 as well as German PPP project of DAAD and the BMBF. 
The financial support by the Brazilian Ministry of Science and Technology is acknowledged.
  
\bibliography{bound_GL}
\end{document}